\documentclass[11pt,preprint]{aastex}
\newcommand{\etal}{{\it et~al.$\;$}}
\newcommand{\ie}{$i.e.,\;$}
\newcommand{\eg}{$e.g.,\;$}
\newcommand{\viz}{$viz.,\;$}
\newcommand{\cf}{$cf.,\;$}
\newcommand{\abt}{$\sim$}

\newcommand{\asec}{$^{\prime\prime}$}
\newcommand{\ang}{$\rm \AA\;$} 
\renewcommand{\deg}{$^\circ$}

\newcommand{\htwo}{H$_2$}
\newcommand{\Lb}{{Ly$\beta\;$}}
\newcommand{\Ha}{{H$\alpha\;$}}
\newcommand{\Hb}{{H$\beta\;$}}

\newcommand{\Pb}{{Pa$\beta\;$}}

\newcommand{\hi}{H~{\small I\,}}
\newcommand{\oiii}{[O~{\small III}]$\lambda5007\;$}

\newcommand{\siii}{[S~{\small III]}$\lambda9069\;$}
\newcommand{\ovi}{O~{\small VI\,}}
\newcommand{\ciii}{C~{\small III\,}}

\newcommand{\kms}{km~s$^{-1}$}

\newcommand{\sn}{$\frac{S}{N}$}
\newcommand{\fuse}{$FUSE\;$}
\newcommand{\iue}{$IUE\,$}
\newcommand{\hst}{$HST\;$}
\newcommand{\hststis}{$HST~STIS\;$}
\newcommand{\stis}{$STIS\;$}

\slugcomment{First submitted 7 Apr05; revised version Sep 05}

\shorttitle {\ovi in an Obscured Seyfert}
\shortauthors{Shastri et al.}

\accepted{2006 March 28}
\begin{document}

\title{\ovi\, Asymmetry and an Accelerated Outflow in an Obscured Seyfert:\\
\fuse and \hststis Spectroscopy of Markarian~533}

\author{Prajval Shastri}
\affil{Indian Institute of Astrophysics, Bangalore, India 560 034}
              \email{pshastri@iiap.res.in}

\author{John Hutchings}
\affil{NRC Herzberg Institute of Astrophysics, 5071 West Saanich Road,
Victoria, BC, V9E 2E7,  Canada}

\author{Jayant Murthy}
\affil{Indian Institute of Astrophysics, Bangalore, India 560 034}

\author{Mark Whittle}
\affil{Dept. of Astronomy, University of Virginia, Charlottesville, VA 22903, USA}

\and

\author{Beverley J. Wills}
\affil{University of Texas Dept. of Astronomy, 
1 University Station, C1400, Austin, TX 78712,  USA}

\begin{abstract}

We present far-ultraviolet spectra of the Seyfert~2 galaxy Mrk~533 obtained
with \fuse. These spectra show narrow asymmetrical 
\ovi$\lambda\lambda$1032,1038 emission lines with stronger wings shortward 
of the peak wavelength, but the degree of asymmetry of these wings in velocity 
is much lower than that of the wings of the lines of lower ionization.  In the combined \ovi profile there are marginal  
indications of local absorptions in the outflow. The \ciii$\lambda$977 line is seen weakly 
with a similar profile, but with very low signal to noise.  These $FUV$ spectra are 
among the first for a Seyfert of type~2, \ie a purportedly obscured Seyfert. 
The \hststis spectral image of Mrk~533 allows delineation of the various 
components of the outflow, and we infer that the outflow is accelerated. 
We discuss the results in terms of nuclear geometry and kinematics.
\end{abstract}


%

\section{INTRODUCTION}

Seyfert (Sy) galaxies have been long-recognised as a class of active galactic 
nuclei (AGN) that occupy the low end of the AGN luminosity range. 
They are primarily characterized by strong optical emission lines 
with implied Doppler broadening
\raisebox{-1mm}{$\stackrel{>}{\small{\sim}}$}~300~km~s$^{-1}$,  
and a wide range of ionization \citep[\eg][]{Osterbrock89}. They 
are also characterized by  strong X-ray emission \citep[\eg][]{Elvis78}. 
In addition, Seyferts are usually radio-quiet, are in
spiral host galaxies, and have central luminosities below the commonly
adopted limit for quasars of 
{$M_B^{total}=-23$} \citep[\eg][]{SchmidtGreen83}.

Seyferts have traditionally been classified into types 1 and 2, 
distinguished primarily by the clear presence or absence of 
a component to the permitted lines in their spectra that is significantly broader 
than their  forbidden lines. The Unified Scheme 
(\eg Antonucci, 1993) attempts to unify the two Seyfert types by hypothesizing 
a ubiquitous obscuring torus around their nuclei;  the central emission
(including the broad permitted line component) is then obscured
when the Seyfert is oriented with its torus edge-on, and it appears to us as a
Sy~2.  When the torus is pole-on, the broad emission lines from the central 
clouds are directly visible, and the Seyfert is classified as of type 1.
The Unification Scheme  found particularly strong 
support in the discovery of  weak polarized broad lines  from Sy~2s 
\citep{AntonucciMiller85,Moran00}, which are interpreted as 
originating from the hidden clouds but scattered  periscopically into our 
line of sight.  

The Unification Scheme is conceptually appealing and many observational 
results are consistent with its predictions, including, 
(i) the detection in Sy~2s of broad lines of \Pb, to which the 
torus is expected to be transparent \citep{Veilleux97}, (ii) the paucity of 
measured ionizing photons in Sy~2s, consistent with  the torus 
attenuating the ionizing continuum emitted along our line of sight 
\citep[\eg][]{Kinney91}, 
(iii) the ``double-cone" morphology of the narrow emission line structures, 
interpreted as due to the torus shadowing the central ionizing photons 
\citep{Schmitt03II, Ruiz05}
and (iv) the similarity of the circumnuclear environments 
\citep{PoggeMartini02} and the luminosities of the molecular 
gas in Sy~1s and 2s \citep{Curran01}. The universal validity of the 
Unification Scheme is still open 
to debate, however, with some results inconsistent with it, including (i) 
Sy~1 host galaxies being of earlier Hubble-type than Sy~2s \citep{Malkan98} 
and  (ii) the galaxy densities around Sy~1s and 2s being dissimilar 
\citep{Dultzin-Hacyan99}.

In the context of the Unified Scheme, we aimed to compare the hot gaseous 
outflows in Seyferts of the purportedly pole-on and edge-on kinds, as 
manifested in the \ovi$\lambda\lambda$1032,1038 emission line doublet 
along with any absorption that might be associated with it. 
This is possible using  \fuse 
\citep[the Far-Ultraviolet Spectroscopic  Explorer:][]{Moos00,Sahnow00}
 that can observe these wavelengths with spectral resolution 
$\lambda/\Delta\lambda$  of up to \abt20,000.  In the 
framework of unification, we can interpret any similarities and differences 
in terms of the geometry and viewing angle.

Here we present the results from 
our first observations, which were of Mrk~533 (NGC~7674),  
a Seyfert~2 at a redshift of 0.0289. It is the brightest galaxy in the  
gravitationally interacting 
compact group Hickson~96 \citep{Verdes-Montenegro97}, and has its host 
galaxy oriented close to face-on with the ratio of major to minor
 axes of \abt~0.91 \citep[RC3:][]{RC3}. Mrk~533 has radio jets on the 
100-pc scale \citep{Momjian03}. 
Spectropolarimetric observations \citep{MillerGoodrich90,Tran95b} 
have revealed a broad (FWHM~\abt~2830~km~s$^{-1}$) \Hb line in polarized light, 
suggesting the presence of a genuine broad line
region  hidden by obscuring material in our line of sight.
Mrk~533 has its optical continuum polarized (probably due to dust 
scattering) at a PA perpendicular to its radio axis  \citep{Tran95b}. It 
also shows
evidence for Compton-thick X-ray emission and a reflection-dominated 
hard X-ray spectrum \citep{Malaguti98}. Further, it shows 
\hi\,  absorption in the nuclear region \citep{Beswick02}, and has  
been long known to have a spectacular blue-wing to its \oiii 
profile \citep[\eg][]{Afanasev80,Unger88}. The physical conditions of its
narrow-line region were modeled from optical and \iue spectra by
\citet{Kraemer94}.

We assume a Hubble constant of 75~\kms~Mpc$^{-1}$ which gives a scale of 
541pc/arcsec.

\section{THE OBSERVATIONS AND DATA REDUCTIONS}

Mrk~533 was observed with the twin Rowland-circle spectrographs on \fuse 
during two separate pointings, 
on 2002~August~18 for 10,478s and on 2002~August~20 for 10,851s.  
The spectra were taken through the largest aperture, which has a size 
30\asec$\times$30\asec\, (the ``LWRS"). The ultraviolet photons are detected on 
micro-channel plate detectors, simultaneously via four independent optical channels, 
each of bandwidth \abt 200\ang. Two of these channels use Al+LiF 
coatings on the optics and the other two use  SiC coatings 
\citep[\cf][]{Sahnow00}. The Mrk~533 data were recorded in a 
total of 10 exposures, in photon-counting ``time-tag" mode. 
Due to astigmatic effects on the spectra, no reliable spatial information 
in the cross-dispersion direction can be obtained. The final spectral 
resolution is primarily constrained by the drifts of the object within the 
aperture. 
Only one of the channels, \viz the LiF1A, is continuously aligned on target by 
the guiding camera (the Fine Error Sensor), whereas the alignment with the 
rest of the channels is done on orbital time-scales. 

Mrk~533 was also observed by \hststis on  2000~September~12. 
The 52\asec$\times$0.2\asec slit was centered
on the peak of the stellar continuum and oriented at a 
position angle (PA) of 124\deg, chosen
to lie close to the two slightly different radio axes
seen in the 15~GHz image of \citet{Unger88}.  We present data obtained
with an 1140~sec exposure using the G430M grating which gives a coverage 
of 280\ang  centered on a wavelength of 5095\ang and
sampled at 0.276\ang/pixel (FWHM=~2.5~pixels). The spatial scale is
0\asec.05/pixel.

\subsection{The Pipelined Data from \fuse}
\label{pipeline}

Pre-processed data from the \fuse pipeline 
\citep[{\it CALFUSE} version 3.1.0,][]{Dixon02} 
were used for this work. 
This pipeline-processing included (a) removing the effects of 
the thermally induced 
drifts of the target image on the detector via modelling;
(b) flat-fielding using the pre-launch two-dimensional flat field and 
the one-dimensional on-orbit flat field obtained by observing 
FUV-bright dwarfs; 
(c) conversion of wavelengths to the heliocentric reference frame and 
their calibration; 
(d) averaging the two-dimensional spectral image in the cross-dispersion 
direction to give a one-dimensional spectrum. 

The relative motion of the spectrum on the detectors 
on an orbital time-scale results in shifting of the zero-point of the 
wavelength scale on the detector. From interstellar features, we found that 
these changes were $\le$~0.15\ang, which at the observed wavelength of the 
\ovi\, line corresponds to $\le$45~\kms. 

\subsection{The Co-added \fuse Spectrum}

Data from all the 10 exposures in each of  the channels were co-added. 
Because of their low effective area and the low signal, the data from the 
SiC channels were not usable, and therefore we only 
used data from the LiF1 and LiF2 channels. Their effective area amounts to 
\abt 26~cm$^2$. Photons from each of these channels are imaged by the 
optics onto two segments of the detector, labelled A and B. Thus the 
LiF1A and LiF2B data have overlapping coverage of  the \abt1000--1090\ang 
region and the LiF2A and LiF1B data both cover the \abt1090--1190\ang region.
A comparison of the count-rates with time in the LiF2
indicates that  relative to the LiF1 channel there is no 
significant systematic loss of signal in it, and therefore  
no significant misalignment of the channels has occured during the exposures. 
The signal in the two LiF channels
is comparable and they were combined to form a single
mean spectrum.

The co-added data were averaged along the cross-dispersion direction to 
obtain a one-dimensional spectrum. 
The pixels were binned  along the dispersion direction by 
a factor of 8 to \abt~0.05\ang per bin. 

\subsection{Derivation of the \fuse Background}

The \fuse background is due to intrinsic detector background, the detector
environment and scattered light. 
We did not use the background subtraction from the {\it CALFUSE} pipeline, 
since  the pipeline extractions gave negative flux values for our data which 
indicates oversubtraction of the background. Instead,  we used 
an empirical subtraction procedure of \citet{MurthySahnow04} which uses the 
neighbourhood pixels next to the spectrum.

The procedure assumes that the value of the 
instrumental background within the aperture is identical to the value 
just off the aperture. 
The raw \fuse data were processed 
through the {\it CALFUSE} pipeline with the 
background-subtraction turned off. The resultant co-added 2-D data 
from each of the two exposures were then used. Data on either 
side of the desired aperture were averaged in the
cross-dispersion direction and smoothed by a box-average in the
dispersion direction to obtain the background spectrum.
Although this procedure results in some smoothing of any wavelength-dependence 
of the background,  it is clearly more representative, and moreover, gives 
flux values that lie above zero.  We note that the spectral dependence 
of the background does not vary between orbital day and orbital night. 
Therefore the subtraction was done without separating the orbital day-  
and night-data.
Further,   within the noise, the background-subtracted data are consistent 
between the two different \fuse exposures, and also between the two LiF 
detectors with overlapping wavelength coverage, which vindicates our procedure.

\subsection{Contamination of the \fuse\, Spectrum by Galactic Molecular \htwo} 
\label{h2model}

As expected, the largest contaminant of the AGN spectra in the \fuse band is 
absorption by molecular \htwo\,  in our Galaxy. We take this 
contamination into account by comparing the dips in our observed spectra 
with simple models of 
absorption by molecular hydrogen at zero redshift \citep{Fullerton01}.
The \htwo\,  absorption models are characterized by a 
single component at a kinetic temperature of 300~K, with zero metallicity, 
the rotational states chosen, a fixed atomic Hydrogen column density 
of $5.0\times10^{20}$cm$^{-2}$ 
and a varying molecular Hydrogen column density.  
More details are discussed in section \ref{spectra}.

\subsection{Reduction of the \hststis Data}

The \hststis data were calibrated within $CALSTIS$ and cleaned using
$IRAF/COSMICRAYS$. The slit location was verified by comparing the
integrated flux with $WFPC2$ observations taken at similar wavelengths.

We note that the position of the \stis slit along the wavelength axis 
cannot be determined with the precision we need in order to calibrate 
the zero point of the wavelength scale. We therefore use the results 
from the data of \citet{Unger88} taken with the IDS on the Isaac-Newton 
Telescope for this calibration. In their spectra, the wavelength of the peak of the \oiii 
emission, and of the highly blueshifted component are 5153.0\ang and 
5134.6\ang respectively, whereas the \stis data with $CALSTIS$ calibration 
give 5155.4\ang and 5137.1\ang respectively, implying an offset between 
the \stis and groundbased data of +2.4\ang and +2.5\ang respectively. 
We therefore applied a shift of $-2.5$\ang to the \stis data after 
calibration with $CALSTIS$.

\section{THE EMISSION LINES IN THE INTRINSIC FAR-$UV$ SPECTRUM OF MRK~533}
\label{spectra}

\begin{figure}[h!]
\includegraphics[angle=90, width=6in,height=3.5in]{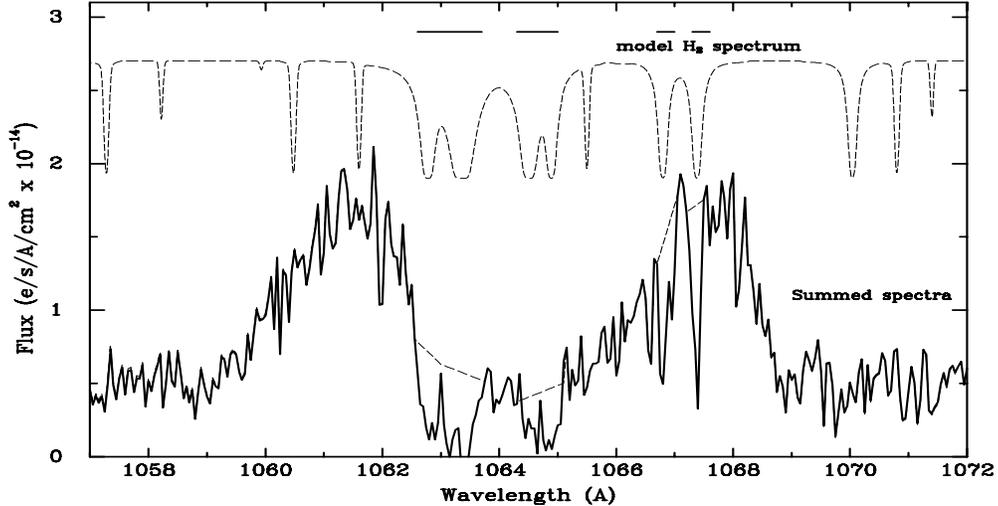}
\caption{The \ovi\, doublet region of the spectrum of Mrk~533.  
Data from the
detectors LiF1A \& LiF2B which give redundant coverage of this region, have been combined (lower panel). 
On top in this panel  an example of a model absorption spectrum of
Galactic \htwo\, at zero redshift that corresponds to a \htwo\, column density 
of 5$\times~10^{19}cm^{-2}$ is illustrated (\cf \ref{h2model}), 
showing that several of the dips in the
Mrk~533 spectra are contamination from absorption in our Galaxy. The positions 
of these troughs are indicated by horizontal lines above the upper \htwo\, spectrum.
The dashed lines marked on the \ovi spectrum indicate a rough removal of the effects 
of this absorption (\cf Sec.~\ref{spectra}). }
\label{OVI}
\end{figure}

\begin{figure}[h!]
\includegraphics[angle=90,width=6in,height=4in]{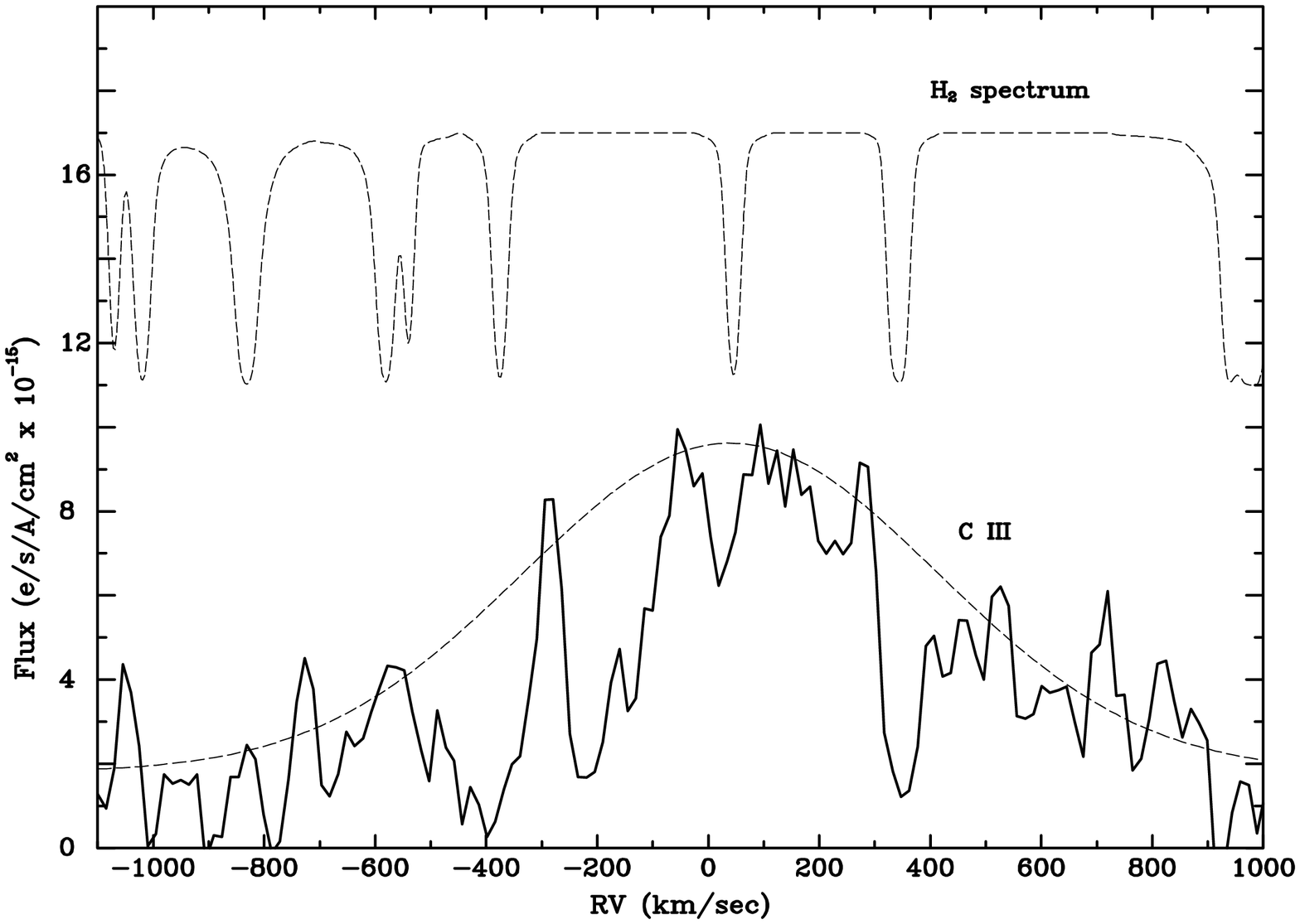}
\caption{The spectrum in the \ciii region plotted in velocity space. 
The S/N is much lower in this region, but nevertheless the \ciii emission is 
detected.  The model \htwo\, absorption spectrum at zero redshift is plotted on 
top.  A symmetrical (Gaussian) profile is plotted to guide the eye, but no attempt 
has been made to remove the effect of Galactic \htwo\, absorption.
A trough is seen (in both detectors, combined here) in  a region 
that is clear of Galactic \htwo\, absorption, which corresponds to 
a velocity of $-$200\kms, but is at best of marginal significance.}
\label{CIII}
\end{figure}

The only strong line feature detected is the \ovi\, doublet
 (Fig.~\ref{OVI}). 
The \ciii emission at 977\ang is seen weakly at its expected 
redshifted position (Fig.~\ref{CIII}).  We do not detect the 
\Lb~$\lambda$1026, the He~{\small II}~$\lambda$1085 or the
Ne~{\small III}~$\lambda$991 lines that were seen \eg in the
{\it HUT} spectrum of the archetypal Sy~2, NGC~1068 \citep{Kriss92}. 
They would have been seen clearly at the equivalent widths observed in NGC~1068, 
as the continuum signal level in Mrk~533 in those  wavelength regions are comparable to
that near the \ovi\, doublet.

Also plotted in Fig.\ref{OVI} is a model used for absorption 
by molecular hydrogen. The corresponding \htwo\, column density is 
$5\times10^{19}$~cm$^{-2}$, although we note that this 
model is not a unique one corresponding to this column density. An inspection of the 
dips in the model and comparison of their widths with the dips in the 
observed spectra clearly indicate that there is a correspondence between 
the strongest \htwo\,  absorptions (marked in Fig.~\ref{OVI}) and some of the dips in 
the observed spectra. The comparison model was chosen  
by inspection to have 
widths of the strongest absorptions comparable with those seen in the 
observed spectrum. The widths of the main absorptions
in the region depend significantly on the column density.
The data are consistent with the observed  
dips being due to the strongest \htwo\, absorptions with a shift 
of $-$0.15\ang. This is within the uncertainty in the calibration of 
the wavelength zero point of the \fuse spectrum.   

\begin{figure}[h!]
\includegraphics[angle=90, width=6in,height=3.5in]{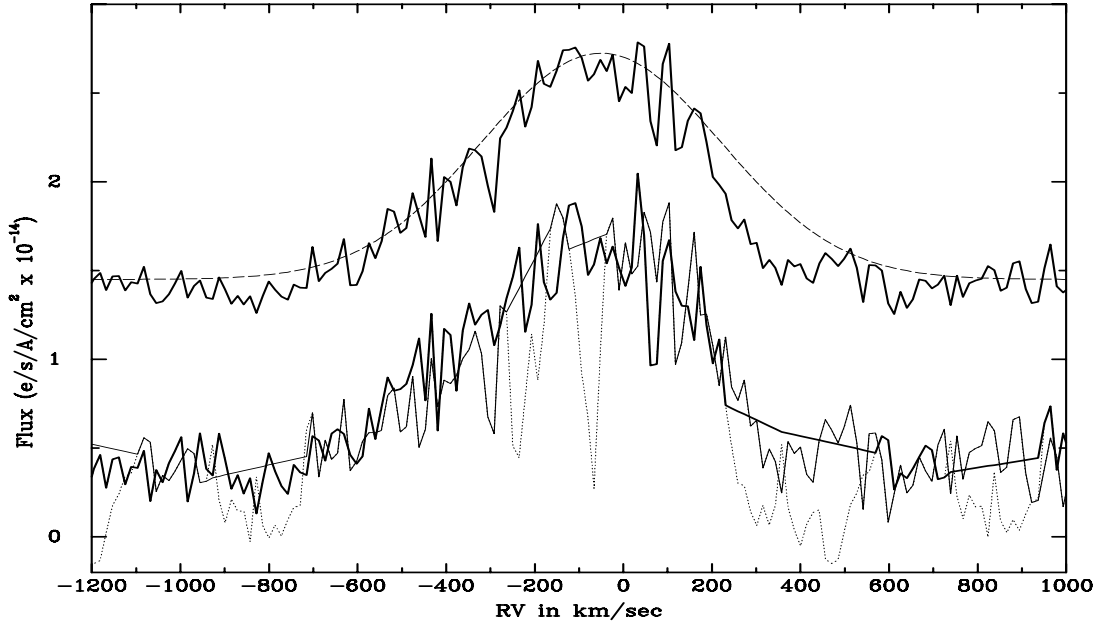}
\caption{ The Mrk~533 spectra in the O~VI region plotted in velocity space.
A systemic redshift of 0.02888 (heliocentric velocity of 8659~\kms) 
and a doublet separation of 5.85\ang has been assumed.
In the lower plot the heavy and light solid lines represent the 
spectra of the 1032\ang and 1038\ang lines respectively. The dotted 
curves represent the dips in these spectra that are assumed to be 
due to \htwo\, absorption, and the corresponding solid lines show their 
rough removal (see \ref{spectra}  for further details).  The averaged profile 
is plotted in the top panel along with a symmetrical (Gaussian) profile  
for comparison, which illustrates the shortward asymmetry of the observed profiles, 
as well as the suggestive residual  absorption troughs at 
\abt$-$300\kms and $-$800\kms. If real, these troughs could be 
 intrinsic to Mrk~533.}
\label{OVIvel}
\end{figure}

Fig.~\ref{OVIvel} we show the profiles of the \ovi doublet 
lines in velocity space. 
There is a small slope to the continuum
around the \ovi\,  lines in the range 1050-1075\ang that rises with wavelength   
(Fig.~\ref{OVI}). This slope is  likely to be real and is also present 
in the archival spectrum that uses the \fuse pipeline background subtraction. 
We discuss possible extinction effects later.
As we are concerned here with the emission line profiles, in plotting 
Fig.~\ref{OVIvel} we have simply subtracted the slope in the continuum across
the \ovi  emission lines,  by subtracting a linear continuum using 
the points at 1057--1058\ang and 1071--1073\ang. The diagrams show a level
continuum at 
$4\times~10^{-15}~ergs~cm^{-2}~s^{-1}~ang^{-1}$.

The heliocentric systemic velocity of 
8659~\kms (see \ref{shifts}) was adopted for the velocity zero 
of the plot.  The shift between the profiles of 5.85\ang is the  
separation of the OVI doublet redshifted by 0.02888. Apart from 
the removal of the linear slope, no other scaling was done.

Clearly the two profiles match within the noise except for 
a few troughs which in fact correspond to those in the illustrated 
\htwo\, absorption model.
We show in Fig.~\ref{OVIvel} the result of a rough removal of this assumed contamination, 
estimated by comparing with the model \htwo\, spectrum. 
The heavy solid and light solid curves represent the 1032\ang and 1038\ang 
lines respectively. 
All the dips in each of the observed profiles  that are assumed 
to be due to contaminating  Galactic \htwo\, absorption are shown as 
dotted. 
In the case of each removal, the straight line 
compares well with the profile of the other line in that 
velocity region within the noise. The removals are also indicated 
in Fig.~\ref{OVI}.

Given the good match of the profiles of the \ovi\, doublet, they were 
combined to form a mean, plotted in Fig.~\ref{OVIvel}. 
The profiles show an asymmetry  about zero velocity in the form of an extension 
toward shorter wavelengths.  Two possible dips remain in the 
average profile, corresponding to velocities of $-$300\kms and $-$800\kms.  

The spectrum of the \ciii$\lambda$977 line (\cf Fig.~\ref{CIII})
is rather noisy, and apart from its detection, no definitive statement 
can be made about its profile. While it is interesting that data from 
both the LiF channels show a relatively deep dip in a region that is clear 
of any \htwo\, absorption and corresponds to a velocity of 
\abt$-200$\kms, we consider this to be  of only marginal significance 
in view of the low \sn.

\section{THE \oiii VELOCITY STRUCTURE OF Mrk~533 FROM \hststis}
\label{specim}

\begin{figure}[ht]
\includegraphics[width=6in]{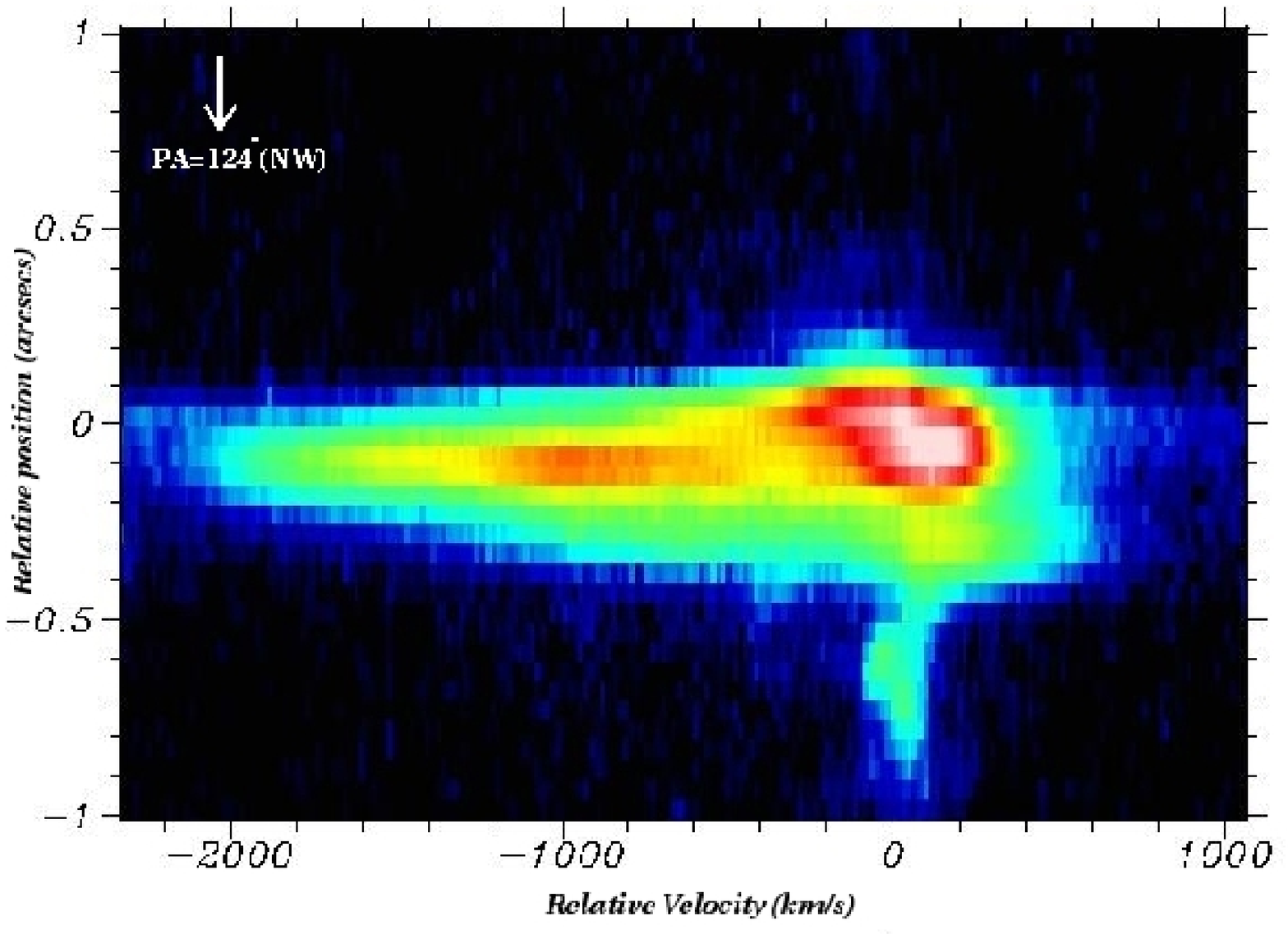}
\caption{The \hststis spectral image of Mrk~533 in the region 
of \oiii line is plotted in velocity space.
The vertical spatial axis is along the \stis slit at a PA of 124\deg 
(along the radio axis).  The zero position of this axis corresponds
to  the peak of the continuum in the off-line region. The zero position of the horizontal 
dispersion axis corresponds to the systemic velocity of 8659\kms\, 
(\cf Sec.\ref{specim}).}
\label{SpecImage}
\end{figure}

In Fig.\ref{SpecImage}, the \hststis spectral image in the rest-frame 
\oiii spectral region is plotted in velocity space. The vertical axis is 
in the direction of the \stis slit which was oriented parallel to the radio 
axis (PA~=~124\deg). The spatial zero point on this axis is the nominal 
nuclear position, taken to be the peak in the continuum. (It is 
possible that the true nucleus is shifted w.r.t. this position 
since the obscuring dust in the line of sight would be expected to attenuate its 
emission, thereby effectively shifting the continuum peak.)
The zero position on the horizontal dispersion axis corresponds to the heliocentric systemic velocity.

Clearly, the \oiii emission is spatially asymmetric, with most of the
emission to the NW of the nominal nuclear position,  
along the radio axis \citep[see images in][]{Unger88, Momjian03}. 
On the continuum peak, the 
\oiii profile peaks close to the systemic velocity, and has a 
``blue shoulder" at \abt$-$100\kms 
\citep[similar to what was reported earlier by ][]{Unger88, Veilleux91III}. 
Roughly 0\asec.1 to the NW, the 
peak shifts to \abt+200\kms\, while there is a broad blueshifted peak
at \abt$-$1000\kms with an extended wing reaching \abt$-$2000\kms.  The spatial
extent of this broad maximum appears to be 0~--~0\asec.2.

\section {DISCUSSION}

We assume in our analysis  that all the spectral emission in the 
\fuse data is attributable
to the AGN  and its photoionization effects, and that even in the 
30\asec$\times$30\asec\, aperture, the contribution from hot stars is negligible. 
This assumption is supported by a comparison of the Mrk~533 spectrum
with the \fuse spectra from hot stars in the atlas of
\citet{Walborn02}. None of the expected strong emission lines from stars is seen in our spectra, so we assume that
all emission arises in the AGN.

We also recall that in \eg NGC~3516,  \citet{Hutchings01} find from a comparison of 
contemporaneous large-aperture \fuse spectra and narrower-aperture \hststis spectra 
that the continumm levels 
in the overlapping spectral region match very well. They conclude therefore, albeit for a Sy~1,  
that the continuum is dominated by nuclear emission and there is no 
significant contribution to the \fuse photons from starlight at larger radii.

\subsection {The Line-Ratio of the \ovi\, Doublet} 

The statistical weights of the upper levels of the \ovi\, doublet transitions are 
two and four for the longward and shortward wavelength lines respectively, implying 
an intensity ratio of two for the doublet under the 
assumptions that the emission is optically thin and that the emitters are in LTE 
\citep{MaucheRaymond98}. In the Sy~1s, this is indeed the typical ratio that 
has been found for the broad emission-line region 
(Kriss, private communication). 

In Mrk~533, however, the two doublet lines of the \ovi\,
emission match very well in both intensity and profile 
as can be seen from Fig.~\ref{OVIvel}. We estimate the 
line flux in both cases to be \abt~3.7$(\pm0.2)\times10^{-14}~ergs~s^{-1}~cm^{-2}$.
Thus the doublet
ratio is \abt~1, similar to the value inferred from {$HUT$} data for NGC~1068
by \citet{Kriss92}  and in contrast to the trend seen in Sy~1s.
This means that the \ovi\, line is optically thick over roughly the 
whole velocity range in Mrk~533. These aspects 
are being investigated in detail in  a subsequent paper  
(P.~Shastri \etal in preparation).  

We note that the effect of the rising slope of the continuum in 
the wavelength range 1055-1075\ang  (\cf~Sec.~\ref{spectra})  has been taken out 
in plotting Fig.~\ref{OVIvel} and thus it clearly does not alter our conclusion
about line ratio of the doublet.

\subsection{Shifts of the Lines Relative to Systemic} 
\label{shifts}

We use a heliocentric systemic velocity $cz$~=~8659~\kms,  
which is the mean of the host galaxy stellar
velocity from \citet{NelsonWhittle95} and the values derived by 
\citet{Unger88} and \citet{Nishiura00} from their \Ha rotation curves.
The corresponding heliocentric redshift is 0.02888. 

Within the spectral resolution constraint of the \fuse 
data, the recession velocities derived from the peaks of the \ovi\, doublet 
and the \ciii line are consistent with the systemic velocity.
Right at the nominal nuclear position, the \oiii peak is also at systemic
(Fig.\ref{SpecImage}),
though a larger aperture would find the strongest near-nuclear \oiii
emission to be redshifted by \abt~200\kms. This shift relative to 
systemic was noted from ground based
spectra by \citet[][\oiii]{Unger88} and \citet[][\siii]{NelsonWhittle95}.
It is likely that such spatial and velocity structure accounts for other 
small redshift differences noted in the literature, such as the peak 
\Ha of 8756~\kms measured by \citet{Veilleux91I}. Two other velocities 
deserve mention. The CO emission is centered essentially at systemic 
\citep[8669~\kms,][]{SandersMirabel85}, as one might expect for
dense gas relatively deep in the galaxy potential. Finally, \hi\, is seen in
slightly blueshifted absorption \citep[8623~\kms,][]{Beswick02}
consistent with a modest outflow on \abt~100~pc scales.
Tidal interactions with the other galaxies of the compact group 
\citep{Verdes-Montenegro97} 
can also create shifts in the overall peak of the neutral gas. 

\subsection{Intrinsic Absorption by Molecular Hydrogen} 

We do not find any  \htwo\, absorption that 
is intrinsic to Mrk~533 (\eg from the putative torus, given that it 
is a Sy~2). Indeed, since the nuclear continuum emission 
is expected to be attenuated due to obscuration, it is not 
surprising that it is too weak to measure any absorption 
against it.

\subsection{Absorption in the Outflow?}

Strong absorptions in the profile of the {\it FUV} emission lines 
are a characteristic feature of the Seyfert~1 galaxies observed 
by \fuse \citep[\eg][]{Kriss05}. In the case of Mrk~533, however, 
the evidence for absorption is less definitive. This is not entirely 
surprising for a Sy~2, given that the nuclear continuum is expected to be 
significantly attenuated, and the outflow is expected to be primarily 
along the AGN axis, and obscured in the near-nuclear region. 

The combined profile in Fig.~\ref{OVIvel}  shows two dips in both lines of the \ovi\, 
doublet, at $-$300~\kms and $-$800~\kms (see Fig.\ref{OVIvel}). 
These are residual dips that remain after accounting for absorption by 
Galactic \htwo, and have no local identification. Their significance 
is not high, however, and better \sn\, data are required for any 
definitive interpretation. If real, they would have to be due to  
absorbers intrinsic to Mrk~533: 
It is possible that the absorption at $-$300~\kms is associated with 
the outflow from the nucleus, but could also be from an interstellar cloud 
associated with the circumnuclear starburst that is known to be 
present in Mrk~533. The latter absorption (at $-$800~\kms) on the other hand has 
too large a velocity to be associated with such a starburst, which would 
imply that it would have to  be due to a foreground cloud associated 
with the outflow.

In the case of the \ciii emission line, our data are very noisy, and 
although there is a suggestion of a dip in both the photon channels 
in a region clear of any Galactic \htwo\, absorption which corresponds 
to a velocity of $-$200~\kms\,, this feature is at best of marginal 
significance.

\subsection{The Line Profiles and their Blue-ward Asymmetry}

\begin{figure}[h!]
\includegraphics[angle=90,width=6in,height=4in]{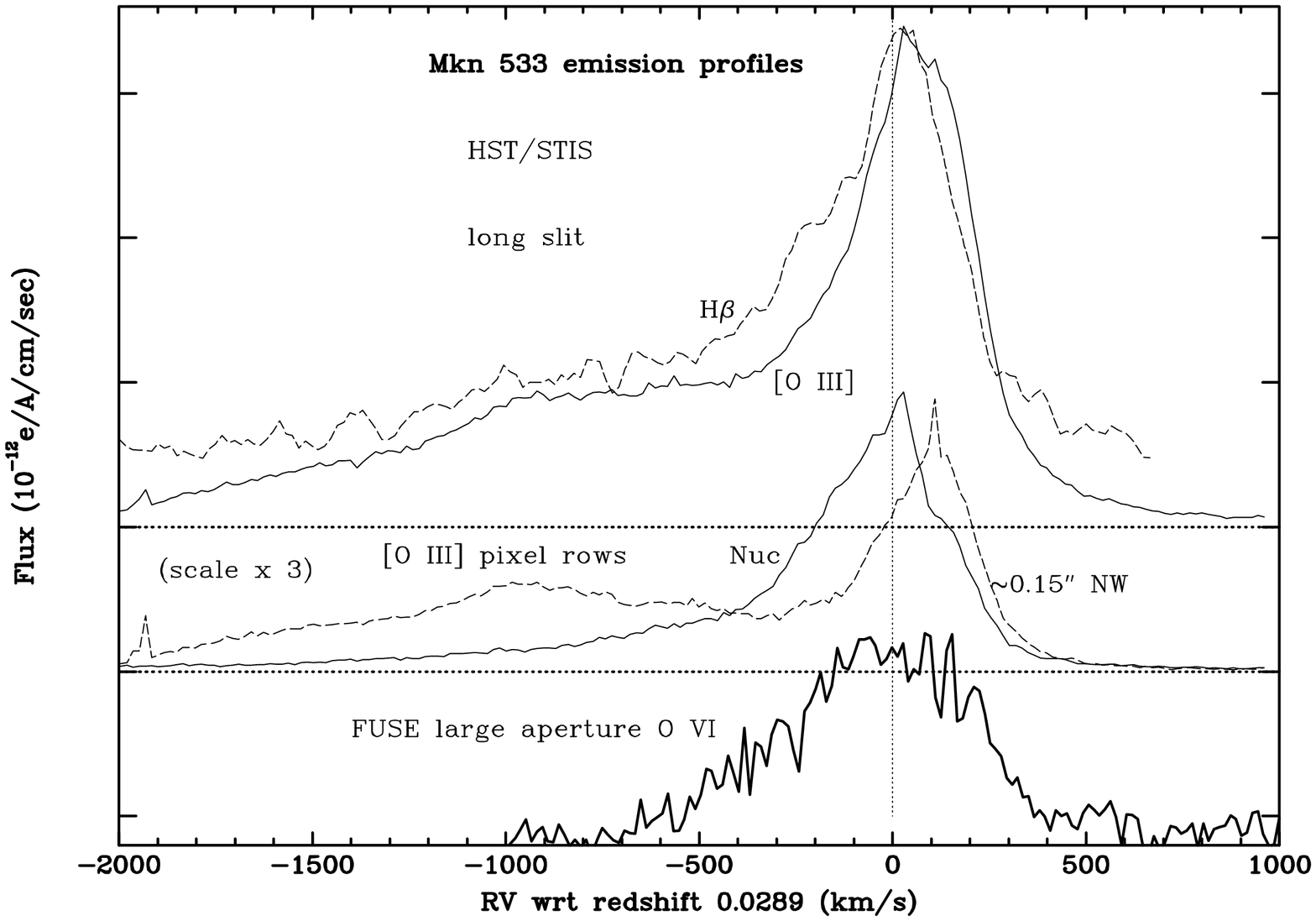}
\caption{The profiles of the \oiii and \Hb lines from the \hststis data 
are plotted together with the \ovi\, profile from \fuse. 
In the top panel, the solid and dashed curves represent 
the profiles 
of the \oiii and \Hb lines respectively, as extracted from a \stis 
long-slit aperture. 
In the middle panel, the solid curve represents the 
profile of the \oiii line extracted from a `nuclear' aperture 
(the 42nd pixel row in the \stis data which is at the continuum position). The dashed 
curve represents the profile from the  off-nuclear region, 
\abt~0\asec.15 NW from the `nucleus' (pixel row 39). 
The y-scale in this panel is magnified by a factor of 3.
The solid curve in the bottom panel is the \ovi\, profile from the \fuse 
spectrum from Fig.~\ref{OVIvel}, 
to compare and contrast with the \stis profile.}
\label{AllProfiles}
\end{figure}

The fact that the \ovi\, emission lines are narrow (\eg comparable
to the width of the core of the \oiii line, see Fig.\ref{AllProfiles}) is, of course, 
consistent with the basic classification of Mrk~553 as a Seyfert~2. 
In Seyfert~1s, the \ovi\, lines are usually much broader although 
some have an additional narrow component \citep{Kriss04}.

Fig.~\ref{AllProfiles}  compares the \ovi\, and \oiii emission profiles. 
The \stis profiles  in this plot were
extracted using (a) a `nuclear' aperture corresponding to one pixel row
centered on the zero position, (b) an `off-nuclear' aperture corresponding to
pixel row \abt~0\arcsec.15 to the NW relative to zero position,
(c) the whole long-slit aperture. The combined profile of the two \ovi 
doublet lines is also shown.

It has been long known that Mrk~533 has a spectacular blue-wing to its \oiii, 
\Hb and \Ha profiles and a less striking wing in the  NII profile
 \citep[\eg][]{Afanasev80,Unger88,Veilleux91I}.
A striking feature of the \fuse profile of the \ovi\, doublet is that 
it also shows an asymmetry in the same sense. 
However, it lacks the extended wing shortward of the peak wavelength seen 
in the the lower ionization lines to $-$2000~\kms, 
and its peak also appears centered near the
systemic velocity, unlike the long-slit (large aperture) \oiii profile (top panel, 
Fig.~\ref{AllProfiles}). 
\citet{deRobertisShaw90} had earlier shown that the asymmetry of the 
emission lines correlate with the ionization potential, for several 
Seyferts including Mrk~533, but its \ovi\, line clearly does not fit this 
correlation.  

Before concluding that the 
differences between the profiles of the \oiii and \ovi lines are real, 
we must consider the impact of different aperture 
sizes on the profile shapes. In fact, the large size of the \fuse 
aperture introduces two potential concerns. The first is that extended  
\ovi\, emission might be included.
However, the low excitation measured outside the nucleus 
\citep[\eg ${[OIII]}/{H_{\beta}}~<~0.5$,][]{Unger88}
suggests there will be little if any off-nuclear \ovi 
contribution. Second, mis-centering the Mrk~533 nucleus in the aperture
could lead to significant wavelength uncertainty. In our case, however, we 
used the values from the LiF1A (guiding) channel which guarantees good 
centering,  and we have determined that the uncertainty is $\le$~45~\kms
(\cf Sec~\ref{pipeline}). Hence, we conclude that the peak of the \ovi emission is
indeed centered on sytemic within the errors, and that it indeed lacks the very 
extended wing shortward of the peak that is seen in the total \oiii profile. 
This is in fact consistent 
with the \ovi\, emission arising within 0\asec.05 of the nucleus 
since this matches the narrower \oiii profile from this 
innermost region (\cf Fig.~\ref{AllProfiles}).
Thus, although the long-slit aperture of 
the \oiii \stis observation (top panel, Fig.~\ref{AllProfiles})
resembles most closely the \fuse aperture in spatial scale, nevertheless, 
the \ovi profile actually closely resembles in both shape and position w.r.t. systemic,
the {\it nuclear} \oiii profile, 
which we know arises within \abt~0.\asec05 or \abt30pc from the continuum source. 
This suggests that the \ovi\, emission arises from this inner \abt~0.\asec05 
region, 
similar to what has been conjectured for the narrow component to the \ovi\, line in
\eg NGC~3516 \citep{Hutchings01}.
We also note that the similarity of the \ovi and the nuclear \oiii profiles 
argue against complex extinction in this region.

Clearly, beyond  0.\asec05 or \abt30pc from the continuum peak, (and
by implication, beyond at least this distance from the putative nucleus),
the \oiii profile is dominated by the blue-wing. If there
is corresponding redshifted emission from
a receding part of the outflow, it must be hidden from view. 

Of relevance here is the fact that the low-ionization lines of Mrk~533   
\citep{Veilleux91I} do show the extended blue-wings upto \abt$-$2000~\kms.
The case of \Hb from our data is shown in Fig.~\ref{AllProfiles}. 
An archival \hst $FOS$ spectrum shows that the 
Mg~{\small II}$\lambda2798\;$ line has a similar profile. 

In this picture, the ionization parameter drops sufficiently quickly 
with distance from the nucleus so that the \ovi\, line is no longer 
generated in the blue~wing 
gas found \abt0\asec.1~--~0\asec.2 (50~--~100pc) off-nucleus. This 
off-nuclear gas, it would seem, has been accelerated either by a nuclear 
wind or the near-nuclear radio-emitting hotspot (seen, \eg example, in the 
images of \citet{Momjian03} which exhibits structures in approximately the
same location).
Evidence for accelerated flows on scales of 0\asec.2 has been 
found in several Seyferts by \citet{Ruiz05}.  
It is perhaps relevant here that there is also direct evidence for 
acceleration in some Sy~1 galaxies \citep[\eg][]{Hutchings01, deKool97}, 
although on a somewhat smaller physical scale.

The blue wing is most commonly interpreted 
as arising from outflow of line-emitting gas with a distributed
source of opacity (dust) which obscures the far side
\citep[\eg][]{Whittle85I, Veilleux91III}.  The alternative --
infalling dusty clouds emitting preferentially from the side facing the
central source -- seems unlikely in this case, not least because some
of the highest velocity blueshifted gas is seen off-nucleus where we
know the gravitational velocities are a factor 10 smaller.
Fig.~\ref{SpecImage} shows 
that most of the \oiii emission comes from a region NW of the nucleus, which
reinforces the idea that the \oiii emission arises from an approaching 
outflow, the receding side of which is hidden.
The torus in the line of sight could provide the opacity required 
to produce the blue wings, 
by obscuring the photons from the redshifted part of the profile, 
which would be those closest to the nucleus but moving away from us.
It perhaps also contributes significantly to the obscuration of the \oiii 
emission from the SE side of the nuclear region, making the \oiii 
structure appear `one-sided'. It may be recalled that the host galaxy of Mrk~533 
is oriented face-on, an unexceptional fact, given the lack of correlation between 
the host galaxy axis and the AGN axis in Seyferts \citep{Schmitt03II}.

Thus if the 'shoulders' shortward of the peak in both the \oiii and \ovi profiles 
originate in the same gas, it means that the  shortward asymmetry of 
the \ovi emission results from a high excitation outflow 
within 0.\asec05 or \abt30pc from the nucleus, and any 
corresponding redshifted emission from 
the receding part of the outflow is hidden from view.

Obscuration of the receding flow implies that dust
is present and may affect what we see of the
approaching flow too. The small rise in the continuum with 
wavelength implies some extinction of the nuclear continuum. 
The redshifted peak of the \oiii line relative to systemic 0\asec.15 NW 
of the nucleus could be understood in this picture.
\citet{Kraemer94} quote a reddening value E(B-V)=0.20 based on the 
line flux ratios,  
but given the uncertainties in the extinction curve in the \fuse 
range, we are unable to quantitatively comment on the consistency 
of our continuum slope with this reddening  value.
Using our line fluxes for the \ovi, and the \Hb line flux from 
\citet{Kraemer94} the line ratio of $\frac{\ovi}{H\beta}$ is \abt~0.6.  
Patchy dust could of course complicate all the profiles, 
and we would need much more detailed evidence to consider that.  
For now, we note particularly that the \ovi profile is very 
similar to the nuclear [O III] profile.
However, given the significant differences in the overall 
shapes of the \ovi and off-nuclear \oiii profiles even in regions where the \sn\, 
is high, it is very unlikely that such differences can arise from 
dust extinction, and therefore must be principally attributed to velocity and 
ionization gradients.

A further conclusion from the absence of an extended shortward wing in the 
\ovi\,  profile
 is that this high velocity gas is likely to be {\it photoionized} by the 
central source, rather than {\it shock-ionized} by a wind or jet. 
In the shock-ionization picture one would expect, at 
least naively, that the highest velocity gas would have the highest 
ionization. That this is not the case in Mrk~533 argues against shock 
ionization.  A more detailed analysis using lower 
ionization lines gave a similar result for Mrk~78 \citep{Whittle05}, but in 
the case of Mrk~533 the large difference in the degree of ionization 
between O\,$^{5+}$ and O\,$^{2+}$ makes for a stronger case.

Finally, we note
that the picture described here is also consistent with the 
spectropolarimetric results of \citet{Tran95b} who found that 
the \oiii profile in polarized light 
was narrower and had no extended blue wing. The region producing the scattered 
light is likely to be highly nuclear (afterall, the polarized spectrum shows 
broad \Hb) and hence it is not surprising that the scattered \oiii profile 
matches our most nuclear 
\stis data: narrower profile, centered on systemic, and with little blue wing; 
indeed similar to the \ovi profile.

If the outflow is in the form of a hollow bi-cone \citep[\eg][]{Ruiz05},
then, depending on the opening angle and the inclination of the line of sight, 
we may see both approaching and receding velocities along the line-of-sight
from the approaching `cone'.  The obscuring torus may then also affect the 
nuclear part of the emission profile from the approaching `cone', and, \eg 
preferentially obscure the approaching velocities. The redshifted peak of 
the \oiii line relative to systemic could be understood in this picture, 
although we do not have enough spatial resolution to model it in detail.

\section{CONCLUSIONS}

We present high resolution \fuse spectra of the permitted ultraviolet
 \ovi$\lambda\lambda$1032,1038 and \ciii$\lambda$977 lines and \hststis 
spectra of the \oiii optical line, of the Seyfert~2 galaxy Mrk~533
(NGC~7674). We find the following:

\begin{enumerate}

\item The \ovi doublet lines are strong and relatively narrow, consistent with
their origin in the narrow-line region.
This is in predicted contrast to the very broad \ovi\, line reported  in  Seyferts of 
type~1 which are purportedly pole-on.

\item  The \ciii~$\lambda$977 emission line is also weakly detected with similar line width, but
the \Lb~$\lambda$1026, He~{\small II}~$\lambda$1085 and 
 Ne~{\small III}~$\lambda$991 emission lines are not detected.

\item The profiles of the two doublet lines of \ovi emission are very well 
matched, implying a line ratio of \abt~1, and thus optically thick emission. This fact is
also consistent with \ovi doublet not originating in a broad-line region which 
typically shows a flux ratio \abt~2.

\item After correcting for Galactic \htwo\,  absorption, we identify 
with somewhat marginal significance two possible intrinsic absorption 
features in the \ovi profile at $-$300~\kms and $-$800~\kms.

\item The \ovi\, velocity profile is centered near the systemic velocity and has a moderate 
shortward asymmetry.  This matches quite well the \oiii profile 
from the innermost pixel ($<$0\asec.05, \ie $<$30pc) of the \stis data.

\item  However, moving only slightly off nucleus 
(0\asec.1--0.\asec2 or 50-100pc), the \oiii 
profile peak shifts to +100~\kms and  the shortward side gains a strong wing
extending to \abt$-2000$~\kms.

\item  The difference between the \ovi and \oiii profiles is consistent with a
picture in which the most nuclear region exhibits velocities \abt~500~\kms
in mild outflow with the receding part hidden from view. However, beyond this 
inner region, the gas is strongly accelerated by a nuclear wind or jet flow. 
The narrow and symmetric profile of the polarized \oiii line is also 
consistent with this picture.

\item The absence of \ovi emission in the highly accelerated 
\oiii-emitting gas argues in support of this gas being 
photoionized rather than shock-ionized.

\end{enumerate}

\begin{acknowledgements}

Help from the \fuse team during the proposal and observation stages, and the 
efforts that have gone into the high quality of their user services are 
gratefully acknowledged.  We thank Suzy Collin for several insightful 
discussions, and the anonymous referee whose comments improved the paper 
very significantly.  This paper is based on observations made with the NASA-CNES-CSA 
Far Ultraviolet Spectroscopic Explorer. \fuse is operated for NASA by the 
Johns Hopkins University under NASA contract NAS5-32985.
The \hststis spectra are based on observations 
obtained at the Space Telescope Science Institute which is operated by the 
Association of Universities for Research in Astronomy, Inc., 
under NASA contract NAS 5-26555.
Support for the \hst proposal GO-8453 was provided by NASA through a grant 
from the Space Telescope Science Institute.
The \fuse research was supported by NASA grant NAG5-13726.
This research has made use of the Canadian Astronomy Data Centre (CADC), NASA's
Astrophysics Data System bibliographic services and the NASA/IPAC Extragalactic
Database (NED) which is operated by the Jet Propulsion Laboratory,
California Institute of Technology, under contract with NASA.

\end{acknowledgements}
\bibliographystyle{apj}
\bibliography{sey,fuse}
\end{document}